\documentclass{article}

\usepackage{amsmath}
\usepackage{amsthm}
\usepackage{amsfonts}
\usepackage{dsfont}
\usepackage{graphs}

\newcommand{\bv}{\vec{\beta}}
\newcommand{\cv}{\vec{c}}
\newcommand{\gv}{\vec{\gamma}}
\newcommand{\Cbg}[1]{C_{#1}(\bv,\gv)}
\newcommand{\Cnbg}{\Cbg{n}}
\newcommand{\Hg}[1]{H_{#1}(\gv)}
\newcommand{\Hng}{\Hg{n}}
\newcommand{\bt}{\tilde{\beta}}
\newcommand{\gt}{\tilde{\gamma}}
\newcommand{\W}{W}
\newcommand{\Wd}{W^\dagger}
\newcommand{\Wb}{\overline{W}}
\newcommand{\Wh}{\hat{W}}
\newcommand{\Dt}{\tilde{D}}
\newcommand{\dt}{\tilde{d}}
\newcommand{\et}{\tilde{e}}

\newcommand{\A}{\mathcal{A}}
\newcommand{\xv}{\vec{x}}
\newcommand{\xh}{\hat{x}}

\newcommand{\diag}{\operatorname{diag}}
\newcommand{\sumkn}{\sum_{k=1}^n}
\newcommand{\twovec}[2]{\begin{pmatrix}#1\\#2\end{pmatrix}}
\newcommand{\twomatrix}[4]{\begin{pmatrix}#1 & #2\\ #3 & #4\end{pmatrix}}
\newcommand{\st}{\,:\,}
\renewcommand{\mid}{\mathds{1}}
\newcommand{\Henon}{\text{H{\'e}non} }
\newcommand{\Rad}{\operatorname{Rad}}
\newcommand{\Pinv}{P^{-1}}
\newcommand{\blockdiag}[2]{\begin{pmatrix} #1 & & \\ & \ddots & \\ & & #2\end{pmatrix}}
\newcommand{\Spec}{\operatorname{Spec}}
\newcommand{\RRplus}{\reals^2_+}
\newcommand{\half}{\frac{1}{2}}
\newcommand{\halfi}{\frac{i}{2}}
\newcommand{\pb}[1]{\left\{#1\right\}}

\newcommand{\para}[1]{\left(#1\right)}
\newcommand{\paraa}[1]{\big(#1\big)}
\newcommand{\parab}[1]{\Big(#1\Big)}

\newcommand{\bracketb}[1]{\Big[#1\Big]}
\newcommand{\bracketc}[1]{\bigg[#1\bigg]}

\newcommand{\reals}{\mathbb{R}}
\newcommand{\complex}{\mathbb{C}}
\newcommand{\integers}{\mathbb{Z}}

\newtheorem{definition}{Definition}
\newtheorem{proposition}{Proposition}
\newtheorem{lemma}{Lemma}
\newtheorem{theorem}{Theorem}
\newtheorem{corollary}{Corollary}

\numberwithin{equation}{section}

\makeatletter
\newcommand{\ps@draft}{%
\renewcommand{\@oddhead}{\hfil\textit{Preliminary version of \today}\hfil}}%
\renewcommand{\@evenhead}{\@oddhead}%
\renewcommand{\@oddfoot}{\hfil\textrm{\thepage}\hfil}
\renewcommand{\@evenfoot}{\@oddfoot}%
\makeatother

\def\title#1{\def\thetitle{#1}}
\def\author#1{\def\theauthor{#1}}
\def\address#1{\def\theaddress{#1}}
\def\email#1{\def\theemail{#1}}
\newcommand{\makelefttitle}{%
\noindent\vspace{20mm}\\%
\hrule height0.5mm\vspace{3mm}
\noindent{\huge \bf\sf\thetitle}\vspace{0mm}\\%
\hrule height0.5mm
\noindent\vspace{5mm}\\%
{\large\theauthor}\vspace{2mm}\\%
{\small\theaddress.%
\\\texttt{\theemail}}\vspace{3mm}}

\newenvironment{leftabstract}
{\begin{center}
\noindent\it\hspace{3mm}\begin{minipage}[t]{115mm}
\noindent\hspace{-4mm}\textbf{\sf\large Abstract}\vspace{1mm}\\}
{\end{minipage}\end{center}\vspace{7mm}}

\title{Representation theory of $C$-algebras for \vspace{1mm}\\ a higher-order class of spheres and tori}
\author{Joakim Arnlind}
\address{Department of Mathematics, Royal Institute of Technology, S-100 44 Stockholm}
\email{jarnlind@math.kth.se}

\begin{document}

\makelefttitle\vspace{5mm}

\begin{leftabstract}
  \noindent We construct $C$-algebras for a class of surfaces that are
  inverse images of certain polynomials of arbitrary degree. By using the
  directed graph associated to a matrix, the representation theory can be
  understood in terms of ``loop'' and ``string'' representations,
  which are closely related to the dynamics of an iterated map in the
  plane. As a particular class of algebras we introduce the ``\Henon
  algebras'', for which the dynamical map is a generalized \Henon map,
  and give an example where irreducible representations of all
  dimensions exist.  
\end{leftabstract}

\section*{Introduction}

\noindent In \cite{abhhs} fuzzy analogues of spheres and tori were
constructed as \mbox{$C$-algebras} whose irreducible representations
were then classified by using a graph method.  In this paper, we
extend those results to a larger class of surfaces defined as inverses
images of certain polynomials of arbitrary degree. It turns out that
the representation theory of these $C$-algebras can again be
understood in terms of loop and string representations. Moreover, we
will show that classifying irreducible representations amounts to
finding periodic orbits and $N$-strings of a dynamical map
$s:\reals^2\to\reals^2$. As an important subclass of $C$-algebras, we
introduce the \Henon algebras, for which the dynamical map $s$ will be
a generalized \Henon map. In the cases we consider, every surface has
a \Henon algebra as its fuzzy analogue. We will also give an example
of a second order \Henon algebra for which irreducible representations
of all dimensions exist for a fixed value of the parameter $\hbar$.

\newpage

\section{The $C$-algebras}

\noindent For $c,\alpha_0,\alpha_1,\ldots,\alpha_n\in\reals$, we will
consider subsets of $\reals^3$ being inverse images of the polynomial
\begin{equation}
  C(\xv) = c+\half z^2+\sum_{k=1}^{n+1}\frac{\alpha_{k-1}}{2k}\paraa{x^2+y^2}^{k},\label{eq:Cx}
\end{equation}
and we shall write $\Sigma=C^{-1}(0)$. When $\Sigma$ is a
compact surface, it will have the topology of a sphere or a torus.
Following \cite{abhhs}, we introduce the Poisson bracket
\begin{equation*}
  \{f,g\}=\nabla C\cdot\paraa{\nabla f\times \nabla g},
\end{equation*}
and calculate
\begin{equation}\label{eq:xyzpb}
  \begin{split}
    &\pb{x,y}=z\\
    &\pb{y,z}=\alpha_0x + x\sum_{k=1}^{n}\alpha_k\paraa{x^2+y^2}^{k}\\
    &\pb{z,x}=\alpha_0y + y\sum_{k=1}^{n}\alpha_k\paraa{x^2+y^2}^{k}.
  \end{split}
\end{equation}
To define the corresponding $C$-algebra, we replace $\pb{\cdot,\cdot}$
with $[\cdot,\cdot]/i\hbar$ and choose a particular ordering of the
r.h.s. in \eqref{eq:xyzpb}. Setting $W=X+iY$ and $V=X-iY$, we will
choose this ordering to be
\begin{align}
  &[X,Y] = i\hbar Z\label{eq:XY}\\
  &[Y,Z] = i\hbar\alpha_0 X+\frac{i\hbar}{2}\sum_{k=1}^n\bracketc{\bt_k\parab{V(VW)^k+(VW)^k W}+
    \gt_k\parab{V(WV)^k+(WV)^kW}}\label{eq:YZ}\\
  &[Z,X] = i\hbar\alpha_0 Y+\frac{i\hbar}{2i}\sum_{k=1}^n\bracketc{\bt_k\parab{(VW)^k W-V(VW)^k}+
    \gt_k\parab{(WV)^kW-V(WV)^k}}\label{eq:ZX}
\end{align}
with $\bt_k+\gt_k=\alpha_k$ for $k=1,2,\ldots,n$. From these
equations, $Z$ can be eliminated and the two remaining equations can
be rewritten entirely in terms of $W$ and $V$. The result appears in
the following definition:
\begin{definition}\label{def:WVdeformation}
  Let $\bv=(\beta_1,\ldots,\beta_n)$ and
  $\gv=(\gamma_1,\ldots,\gamma_n)$ be vectors in $\reals^n$ such that
  at least one of $\beta_n$ and $\gamma_n$ is non-zero, and let
  $\alpha\in\reals$. Define $\Cnbg$ to be the quotient of the free
  algebra $\complex\langle V,W\rangle$ with the two-sided ideal
  generated by the relations
  \begin{align}
    &W^2V=\alpha W + \sum_{k=1}^n\beta_k(VW)^kW + \sum_{k=1}^n\gamma_k(WV)^kW\label{eq:Weq}\\
    &WV^2=\alpha V + \sum_{k=1}^n\beta_kV(VW)^k + \sum_{k=1}^n\gamma_kV(WV)^k.\label{eq:Veq}
  \end{align}
  We say that the algebra $\Cnbg$ has order $n$.
\end{definition}
\noindent To go from \eqref{eq:XY}--\eqref{eq:ZX} to
\eqref{eq:Weq}--\eqref{eq:Veq} we set $\alpha=-2\hbar^2\alpha_0$,
$\beta_1=-2\hbar^2\bt_1-1$, $\gamma_1=-2\hbar^2\gt_1+2$ and
$\beta_k=-2\hbar^2\bt_k$ and $\gamma_k=-2\hbar^2\gt_k$ for $k\geq 2$.

As an important subclass of
algebras, we introduce the \Henon algebras; a name that will later be
justified by its relation to the generalized \Henon map.
\begin{definition}
  Let $\bv,\gv\in\reals^n$ such that $\bv=(b,0,\ldots,0)$ and
  $\gamma_n\neq 0$. Then we call $\Hng=\Cnbg$ a \emph{\Henon algebra} of
  order $n$.
\end{definition}
\noindent Note that since we have the freedom of choosing $\bt_k$ and $\gt_k$,
as long as $\bt_k+\gt_k=\alpha_k$, every surface of the form
\eqref{eq:Cx} has a \Henon algebra as its fuzzy counterpart.

Let us continue by noting a crucial fact about the algebra $\Cnbg$:
\begin{proposition}\label{thm:DDtcommute}
  In $\Cnbg$ it holds that $[WV,VW]=0$.
\end{proposition}

\begin{proof}
  Multiplying \eqref{eq:Weq} from the left with $V$, and
  \eqref{eq:Veq} to the right with $W$, one immediately obtains
  $WV^2W=VW^2V$, which is equivalent to $[WV,VW]=0$.
\end{proof}

\section{Hermitian representations}

We are interested in finding hermitian representations of the algebra
generated by the relations \eqref{eq:XY}--\eqref{eq:ZX}. This is
equivalent to finding representations $\phi$, of $\Cnbg$, such that
$\phi(W)^\dagger=\phi(V)$. Let us therefore, by a slight abuse of
terminology, call such representations of $\Cnbg$ \emph{hermitian}. In
the following, we will often write $W$ instead of $\phi(W)$, when
there is no risk of confusion. Let us first show that any hermitian
representation of $\Cnbg$ can be decomposed into irreducible
representations.

\begin{proposition}\label{thm:completely_reducible}
  Any hermitian representation of $\Cnbg$ is completely reducible.
\end{proposition}

\begin{proof}
  Let $\phi$ be a hermitian representation of $\Cnbg$. Moreover, let
  $\A$ be the subalgebra, of the full matrix-algebra, generated by
  $\phi(W)$ and $\phi(V)$. First we note that since
  $\phi(V)=\phi(W)^\dagger$, the algebra $\A$ is invariant under
  hermitian conjugation, thus given $M\in\A$ we know that
  $M^\dagger\in\A$.

  We prove that $\Rad(\A)$ (the radical of $\A$), i.e. the
  largest nilpotent ideal of $\A$, vanishes, which implies, by the
  Wedderburn-Artin theorem (see, e.g. \cite{ASS06}), that $\phi$ is completely reducible.  Let
  $M\in\Rad(\A)$. Since $\Rad(\A)$ is an ideal it follows that
  $M^\dagger M\in\Rad(\A)$. For a finite-dimensional algebra,
  $\Rad(\A)$ is nilpotent, which in particular implies that there
  exists a positive integer $m$ such that $\paraa{M^\dagger M}^m=0$.
  It follows that $M=0$, hence $\Rad(\A)=0$.
\end{proof}

\noindent In any hermitian representation, $W\Wd$ and $\Wd W$ will be
two commuting hermitian matrices, by Proposition \ref{thm:DDtcommute}.
Therefore, we can always, by a unitary change of coordinates, choose a
basis such that they are diagonal. We write $W\Wd=D$ and $\Wd
W=\Dt$, where
\begin{align*}
  &D = \diag(d_1,\ldots,d_N)\\
  &\Dt = \diag(\dt_1,\ldots,\dt_N)\\
  &d_i,\dt_i\geq 0\text{ for }i=1,\ldots,N.
\end{align*}
For hermitian representations, equation \eqref{eq:Veq} is the
hermitian transpose of equation \eqref{eq:Weq}. 
Hence, finding hermitian representations of $\Cnbg$ is equivalent to solving the equations
\begin{align}
  &WD = \alpha W+\sumkn\bracketb{\beta_k\Dt^k W+\gamma_kD^k W},\label{eq:WDDt}\\
  &D=W\Wd\text{ and }\Dt=\Wd\W.
\end{align}
Together with the obvious relation $DW=W\Dt$, we write out \eqref{eq:WDDt} in components:
\begin{align*}
  &W_{ij}\bracketc{\alpha+\sumkn\parab{\beta_k\dt_i^k+\gamma_k d_i^k}-d_j}=0\\
  &W_{ij}\parab{d_i-\dt_j}=0.
\end{align*}
If $W_{ij}\neq 0$, we find that
\begin{align*}
  &d_j = \alpha+\sumkn\parab{\beta_k\dt_i^k+\gamma_k d_i^k}\\
  &\dt_j= d_i.
\end{align*}
If we define the map $s$ by
\begin{align*}
  s:\twovec{x}{y}\longrightarrow
  \twovec{\displaystyle \alpha+\sumkn\parab{\beta_ky^k+\gamma_k x^k}}{x}
  \equiv\twovec{\alpha+q(y)+p(x)}{x},
\end{align*}
and $\xv_i=(d_i,\dt_i)$, we can write $\xv_j=s(\xv_i)$ when
$W_{ij}\neq 0$.  We call $s$ the \emph{dynamical map of $\Cnbg$}. For
a \Henon algebra the dynamical map becomes
\begin{align*}
  s\st\twovec{x}{y}\longrightarrow 
  \twovec{\alpha+p(x)+by}{x}
\end{align*}
which is usually referred to as the generalized \Henon map.

From these considerations we get a necessary condition relating the
eigenvalues of $D$ and $\Dt$ through the dynamical map $s$ and the
structure of $W$. This observation suggests that one should find a way
to keep track of the non-zero matrix elements of $W$; for this reason,
we introduce the directed graph of a matrix.

\subsection{Graph representations}

\noindent Let $G=(V,E)$ denote a directed graph with vertex set
$V=\{1,2,\ldots,N\}$ and edge set $E\subseteq V\times V$. We say that
$G=(V,E)$ is the \emph{directed graph (digraph) of the $N\times N$
  matrix $W$} if it holds that
\begin{align*}
  (i,j)\in E\Leftrightarrow W_{ij}\neq 0
\end{align*}
for all $i$ and $j$ in $V$.  When $W$ is the matrix of a hermitian
representation of $\Cnbg$, we simply say that $G$ is a representation
of $\Cnbg$. If $G$ is connected, we say that the representation is
connected. In the following, we will call a directed path from a
transmitter to a receiver a \emph{string}, and a directed cycle a
\emph{loop}. It is a trivial fact that any finite digraph has at least
one string or one loop.

What can we say about graphs being representations of $\Cnbg$? In
fact, it turns out that one can classify all representations by
classifying their digraphs, and the fundamental building-blocks will
be strings and loops. Let us now proceed and try to understand the
structure of these graphs.

Let $G$ be a representation of $\Cnbg$. To each vertex $i\in V$, we
assign the vector $\xv_i=(d_i,\dt_i)$.  If there is an edge from $i$
to $j$, then $W_{ij}\neq 0$ and we must have $\xv_j=s(\xv_i)$ by the
the argument in the previous section. Now, assume that the
representation $G$ has a loop on $n$ vertices. Then there exists a
sequence $(i_1,i_2,\ldots,i_n,i_{n+1}=i_1)$ such that
$(i_k,i_{k+1})\in E$ for $k=1,2,\ldots,n$, which implies that
$s^n(\xv_1)=\xv_1$. Thus, the existence of a loop implies the
existence of a period point of the dynamical map.

Next, we shall prove that loops and strings are in fact exclusive
subgraphs of any representation, i.e. the existence of a loop
prohibits the existence of a string. We prove this by showing that a
representation with a loop is strongly connected. For this, we need
the following lemma.

\begin{lemma}\label{thm:transrec}
  Let $G=(V,E)$ be a representation of $\Cnbg$. Then $i\in V$ is a
  transmitter iff $\dt_i=0$, and $i\in V$ is a receiver iff $d_i=0$.
\end{lemma}

\begin{proof}
  Let $W$ be the matrix of a representation of $\Cnbg$, whose digraph
  is $G$. Since $D=W\Wd$ and $\Dt=\Wd W$, we have
  \begin{align*}
    d_i &= \sum_k W_{ik}\Wb_{ik} = \sum_k |W_{ik}|^2\\
    \dt_i &= \sum_k \Wb_{ki}W_{ki} = \sum_k |W_{ki}|^2
  \end{align*}
  and it follows that $d_i=0$ if and only if $W_{ik}=0$ for all $k$,
  i.e. $i$ is a receiver. In the same way $\dt_i=0$ if and only if
  $W_{ki}=0$ for all $k$, i.e. $i$ is a transmitter.
\end{proof}

\begin{proposition}\label{thm:stronglyc}
  Let $G$ be a connected representation of $\Cnbg$ containing a loop. Then $G$
  is strongly connected, i.e. for every pair of vertices $i,j$ there
  exists a directed path from $i$ to $j$.
\end{proposition}

\begin{proof}
  Let $i$ be a vertex in a loop, and define $V_R(i)=\{j\in V\st
  \exists\text{ a dipath from $i$ to $j$}\}$. We first want to prove
  that $V_R(i)=V$ and that no transmitters or receivers exist. Assume
  that there exists at least one vertex $j$ such that $j\notin
  V_R(i)$.  Let us denote the vertices in $V_R(i)$ by $1,\ldots,m$ and
  the vertices in $V_c=V-V_R(i)$ by $m+1,\ldots,N$.  Since, by
  assumption there is no edge \emph{from} a vertex in $V_R(i)$
  \emph{to} a vertex in $V_c$, the matrix $W$ takes the form
  \begin{align*}
    W=
    \begin{pmatrix}
      A & 0 \\
      B & C
    \end{pmatrix}
  \end{align*}
  with $B\neq 0$ since $G$ is assumed to be connected. We calculate $D$ as
  \begin{align*}
    D = W\Wd=
    \begin{pmatrix}
      AA^\dagger & AB^\dagger \\
      BA^\dagger & BB^\dagger+CC^\dagger
    \end{pmatrix}.
  \end{align*}
  Since $D$ is diagonal, we must have $AB^\dagger=0$ and
  $AA^\dagger=\diag(d_1,\ldots,d_m)$. If we can argue that $A$ is
  always invertible, then we get that $B=0$, which contradicts that
  $G$ is connected, and hence, $V_c=\emptyset$. Therefore, let us now
  show that $A$ is invertible by showing that $d_i>0$ for
  $i=1,\ldots,m$. 

  We have assumed that there is a loop in $A$ and that $i$ is a vertex
  in a loop, i.e. we have that $s^n(\xv_i)=\xv_i$, where $n$ is the
  number of vertices in the loop. Assume that there is a transmitter
  (receiver) $j$ in $V_R(i)$. By definition of $V_R(i)$ there is a
  non-negative integer $k$ such that $s^k(\xv_i)=\xv_j$, and we also
  know that there is a vertex $l$ in the loop such that
  $\xv_l=s^k(\xv_i)=\xv_j$. By Lemma \ref{thm:transrec} we conclude
  that $l$ is also a transmitter (receiver), which contradicts that
  $l$ is part of a loop. Hence, there are no transmitters or receivers
  in $A$, which, by Lemma \ref{thm:transrec}, implies that $d_i>0$ for
  $i=1,\ldots,m$. This proves that $A$ is invertible, which implies
  that $B=0$, which contradicts the fact that $G$ is connected. We
  conclude that $V_R(i)=V$ for all vertices $i$ in any loop.

  Finally, let us argue that $V_R(i)=V$ for any $i\in V$.  If we
  follow an outgoing dipath from $i$, we must, in a finite number of
  steps, reach a vertex contained in a loop, since there are no
  transmitters or receivers (and the graph is finite). From this
  vertex, by the argument above, we can reach any other vertex through
  a dipath. Hence, $V_R(i)=V$.
\end{proof}

\noindent Since a strongly connected graph can not have any
transmitters or receivers, we get the following result.

\begin{corollary}\label{thm:loopnostring}
  Let $G=(V,E)$ be a connected representation of $\Cnbg$ containing a
  loop on $n$ vertices. Then $G$ does not contain a string.
\end{corollary}

\noindent Moreover, since either a string or a loop must exist in
a finite directed graph, the following definition is natural.

\begin{definition}
  A \emph{String representation} of $\Cnbg$ is a representation whose
  graph does not contain a loop. A \emph{Loop representation} is a
  representation whose graph does not contain a string.
\end{definition}

\noindent From Corollary \ref{thm:loopnostring}, we
conclude that every connected hermitian representation of $\Cnbg$ is
either a string representation or a loop representation. In general,
any hermitian representation is a direct sum of string and loop
representations.

Let us show that the structure of loop and string representations is
preserved among equivalent representations.

\begin{proposition}
  Let $\phi$ and $\phi'$ be two equivalent hermitian representations
  of $\Cnbg$. If $\phi$ is a loop representation then $\phi'$ is a loop
  representation.
\end{proposition}

\begin{proof}
  Since the representations are equivalent, there exists an invertible
  matrix $P$ such that
  \begin{align*}
    \phi'(W) = P\phi(W)P^{-1}\text{ and }\phi'(W^\dagger)=P\phi(W^\dagger)P^{-1},
  \end{align*}
  from which it follows that 
  \begin{align*}
    \phi'(D) = P\phi(D)\Pinv\text{ and }\phi'(\Dt)=P\phi(\Dt)\Pinv. 
  \end{align*}
  Hence, $\phi'(D)$ and $\phi(D)$ have the same eigenvalues, and the
  same is true for $\phi'(\Dt)$ and $\phi(\Dt)$. By assumption, $\phi$
  is a loop representation, which implies that no eigenvalues of
  $\phi(D)$ or $\phi(\Dt)$ are zero, by Lemma \ref{thm:transrec}.
  Hence, no eigenvalues of $\phi'(D)$ or $\phi'(\Dt)$ are zero, which
  implies that $\phi'$ is a loop representation.
\end{proof}

\section{The structure of locally injective representations}

Let us introduce the spectrum of a representation.

\begin{definition}
  Let $\phi$ be a $N$-dimensional hermitian representation of $\Cnbg$. We set
  \begin{align*}
    \Spec(\phi)=\{\xv_i=(d_i,\dt_i)\st i=1,2,\ldots,N\}
  \end{align*}
  and we call this set the \emph{spectrum of $\phi$}.
\end{definition}

\noindent We will now show that the spectrum is preserved among
equivalent representations. This follows from the next lemma.

\begin{lemma}\label{lemma:equiv_permutation}
  Let $D$ and $\Dt$ be diagonal matrices and assume that there exists
  an invertible matrix $P$ such that $PD\Pinv$ and $P\Dt\Pinv$ are
  diagonal. Then there exists a permutation $\sigma$ such that
  $PD\Pinv=\sigma^\dagger D\sigma$ and $P\Dt\Pinv=\sigma^\dagger
  \Dt\sigma$.
\end{lemma}

\begin{proof}
  By an overall permutation, we can write $D$ and $\Dt$ in the
  following block-diagonal form:
  \begin{align*}
    D = \blockdiag{d_1\mid_{n_1}}{d_k\mid_{n_k}}\text{ and }
    \Dt = \blockdiag{\Dt_1}{\Dt_k}
  \end{align*}
  with $d_i\neq d_j$ whenever $i\neq j$. Since $PD\Pinv$ is diagonal,
  it has the same eigenvalues as $D$, including multiplicities.
  Therefore, there exists a permutation $\sigma_0$ such that
  $PD\Pinv=\sigma_0^\dagger D\sigma_0$. From this it follows that
  $[\sigma_0 P,D]=0$ which imposes the following form of $\sigma_0 P$:
  \begin{align*}
    \sigma_0 P = \blockdiag{P_1}{P_k}.
  \end{align*}
  Since $P\Dt\Pinv$ is diagonal, $\sigma_0 P\Dt\Pinv\sigma_0^\dagger$ will also be diagonal. On the other hand
  \begin{align*}
    \sigma_0 P\Dt\Pinv\sigma_0^\dagger =
    \blockdiag{P_1\Dt_1\Pinv_1}{P_k\Dt_k\Pinv_k},
  \end{align*}
  which implies that $P_i\Dt_i\Pinv_i$ is diagonal for $i=1,\ldots,k$.
  Hence, there exists permutations $\gamma_1,\ldots,\gamma_k$ such
  that $P_i\Dt_i\Pinv_i=\gamma_i^\dagger\Dt_i\gamma_i$. Now, let us set
  \begin{align*}
    \gamma=\blockdiag{\gamma_1}{\gamma_k}
  \end{align*}
  and define $\sigma=\gamma\sigma_0$. We then get
  \begin{align*}
    \sigma^\dagger D\sigma = \sigma_0^\dagger\gamma^\dagger D\gamma\sigma_0=
    \sigma_0^\dagger D\sigma_0=PD\Pinv,
  \end{align*}
  and
  \begin{equation*}
    \begin{split}
    P\Dt\Pinv &= \sigma_0^\dagger\blockdiag{P_1\Dt_1\Pinv_1}{P_k\Dt_k\Pinv_k}\sigma_0
    = \sigma^\dagger\gamma\blockdiag{\gamma_1^\dagger\Dt_1\gamma_1}{\gamma_k^\dagger\Dt_k\gamma_k}\gamma^\dagger\sigma\\
    &=\sigma^\dagger\Dt\sigma,
    \end{split}
  \end{equation*}
  since $\sigma_0=\gamma^\dagger\sigma$ and $P_i\Dt_i\Pinv_i=\gamma_i^\dagger\Dt_i\gamma_i$.
\end{proof}

\noindent We will now introduce the concept of \emph{locally injective
  representations}. It is a technical condition that is needed as an
assertion in Theorem \ref{thm:solution}.

\begin{definition}
  Let 
  \begin{align*}
    s:\twovec{x}{y}\longrightarrow
    \twovec{\displaystyle \alpha+\sumkn\parab{\beta_ky^k+\gamma_k x^k}}{x}
  \end{align*}
  be the dynamical map of $\Cnbg$, and let $\phi$ be a hermitian representation of $\Cnbg$. If 
  \begin{align*}
    s|_{\Spec(\phi)}\st\Spec(\phi)\rightarrow\reals^2
  \end{align*}
  is injective, we say that $\phi$ is a \emph{locally injective representation.}
\end{definition}

\noindent Note that if $s$ is invertible, then any representation is
locally injective. In particular, this is true for the \Henon
algebras. It also turns out to be true for all loop representations.

\begin{proposition}
  Let $G$ be a connected loop representation of $\Cnbg$. Then $G$ is
  locally injective.
\end{proposition}

\begin{proof}
  Assume that $s(\xv_i)=s(\xv_j)$ for some vertices $i$ and $j$. Since
  $G$ is a loop representation, there exists a positive integer $n$ such that
  $s^n(\xv_i)=\xv_i$. From Proposition \ref{thm:stronglyc} we know
  that there exists a non-negative integer $k$ such that
  $s^k(\xv_i)=\xv_j$. Since $s(\xv_i)=s(\xv_j)$ we get that
  \begin{equation*}
    \xv_i=s^n(\xv_i)=s^n(\xv_j)=s^{n+k}(\xv_i)=s^k(\xv_i)=\xv_j.\qedhere
  \end{equation*}
\end{proof}

\noindent Now, let us prove the main theorem, giving the structure of locally
injective representations. It enables us to show that every
representation is a direct sum of loops and strings.

\begin{theorem}\label{thm:solution}
  Let $\phi$ be an $N$-dimensional connected locally injective
  hermitian representation of $\Cnbg$.  Then there exists a positive
  integer $k$ dividing $N$, a unitary $N\times N$ matrix $T$, unitary
  $N/k\times N/k$ matrices $U_0,\ldots,U_{k-1}$ and
  $x_0,y_0,\et_0,\ldots,\et_{k-1}\in\reals$ such that
  \begin{align}
    &T\phi(W)T^\dagger = 
    \begin{pmatrix}
      0               &  \sqrt{\et_1}\,U_1  & 0                & \cdots & 0 \\
      0               &  0                & \sqrt{\et_2}\,U_2  & \cdots & 0 \\
      \vdots          & \vdots            & \ddots           & \ddots & \vdots\\
      0               & 0                 & \cdots           & 0 &  \sqrt{\et_{k-1}}\,U_{k-1} \\
      \sqrt{\et_0}\,U_0 & 0                 & \cdots           & 0 & 0
    \end{pmatrix}\\
    &\et_l = s^l(x_0,y_0)\cdot\twovec{0}{1},
  \end{align}
  with $\et_1,\ldots,\et_{k-1}>0$.
\end{theorem}

\begin{proof}
  Let $U$ be a unitary matrix such that $UDU^\dagger$ and $U\Dt
  U^\dagger$ are diagonal, set $\Wh=U\phi(W)U^\dagger$ and let $G$ be
  the digraph of $\Wh$. Let $\xh_0,\ldots,\xh_{k-1}$ be an enumeration
  of $\Spec(\phi)$ such that $\xh_{i+1}=s\para{\xh_i}$ for
  $i=0,\ldots,k-2$. This can be done since $G$ is connected and
  locally injective, which in particular means that if
  $s(\xv_i)=s(\xv_j)$ then $\xv_i=\xv_j$. Moreover, let us write
  $\xh_i=(e_i,\et_i)$. We note that if $G$ has a transmitter, it must
  necessarily correspond to the vector $\xh_0$, in which case
  $\et_0=0$. In particular this means that no vertex corresponding to
  $\xh_i$, for $i>0$, can be a transmitter and hence, by Lemma
  \ref{thm:transrec}, $\et_1,\ldots,\et_{k-1}>0$. Now, define
  \begin{align*}
    V_i = \{j\in V\st \xv_j=\xh_i\}\qquad i=0,\ldots,k-1,
  \end{align*}
  and set $l_i=|V_i|$. Since $\xh_{i+1}=s(\xh_i)$ and $\phi$ is
  locally injective, a necessary condition for $(i,j)\in E$ is that
  $j=i+1$. This implies that there exists a permutation $\sigma\in
  S_N$ (permuting vertices to give the order $V_0,\ldots,V_{k-1}$)
  such that
  \begin{align*}
    W' := \sigma \Wh\sigma^\dagger = 
    \begin{pmatrix}
      0               &  W_1  & 0                & \cdots & 0 \\
      0               &  0                & W_2  & \cdots & 0 \\
      \vdots          & \vdots            & \ddots           & \ddots & \vdots\\
      0               & 0                 & \cdots           & 0 &  W_{k-1} \\
      W_0 & 0                 & \cdots           & 0 & 0
    \end{pmatrix}    
  \end{align*}
  In this basis we get
  \begin{align*}
    D &= \diag(\underbrace{e_0,\ldots,e_0}_{l_0},\ldots,\underbrace{e_{k-1},\ldots,e_{k-1}}_{l_{k-1}})=
    W'W'^\dagger=\diag(W_1W_1^\dagger,\ldots,W_{k-1}W_{k-1}^\dagger,W_0W_0^\dagger)\\
    \Dt &= \diag(\underbrace{\et_0,\ldots,\et_0}_{l_0},\ldots,\underbrace{\et_{k-1},\ldots,\et_{k-1}}_{l_{k-1}})= 
    W'^\dagger W'=\diag(W_0^\dagger W_0,W_1^\dagger W_1,\ldots,W_{k-1}^\dagger W_{k-1}),
  \end{align*}
  which gives $W'_i W'^\dagger_i=e_{i-1}\mid_{l_{i-1}}$ and $\W'^\dagger_i W'_i =
  \et_i\mid_{l_i}$.  Since $\xh_{i+1}=s(\xh_i)$ we know that
  $\et_{i+1}=e_i$, which implies that $W'_i\W'^\dagger_i=\et_i\mid_{i-1}$ for
  $i=1,\ldots,k-1$. Any matrix satisfying such conditions must be a
  square matrix, i.e.  $l_i=l_{i-1}$ for $i=1,\ldots,k-1$. Hence, $W'_i$
  is a square matrix of dimension $N/k$, and there exists a
  unitary matrix $U_i$ such that $W'_i=\sqrt{\et_i}U_i$. Moreover, we
  take $T$ to be the unitary matrix $\sigma U$.  
\end{proof}

\noindent Having obtained this result, we can proceed to the task of
classifying all representations up to equivalence. We start by proving
the following simple lemma:

\begin{lemma}\label{thm:equiv_sum}
  Let $W_1$ and $W_2$ be matrices such that
  \begin{align*}
    W_1 = 
    \begin{pmatrix}
      0               &  w_1U_1  & 0                & \cdots & 0 \\
      0               &  0                & w_2U_2  & \cdots & 0 \\
      \vdots          & \vdots            & \ddots           & \ddots & \vdots\\
      0               & 0                 & \cdots           & 0 &  w_{n-1}U_{n-1} \\
      w_0U_0          & 0                 & \cdots           & 0 & 0
    \end{pmatrix}
    ;\,\,W_2 = 
    \begin{pmatrix}
      0               &  w_1\mid          & 0                & \cdots & 0 \\
      0               &  0                & w_2\mid          & \cdots & 0 \\
      \vdots          & \vdots            & \ddots           & \ddots & \vdots\\
      0               & 0                 & \cdots           & 0 &  w_{n-1}\mid \\
      w_0V            & 0                 & \cdots           & 0 & 0
    \end{pmatrix}
  \end{align*}
  where $U_0,\ldots,U_{n-1}$ are unitary matrices,
  $w_0,\ldots,w_{n-1}\in\complex$ and $V$ a diagonal
  matrix such that
  \begin{align*}
    SVS^\dagger = U_1U_2\cdots U_{n-1}U_0
  \end{align*}
  for some unitary matrix $S$. Then there exists a unitary matrix $P$
  such that
  \begin{align*}
    &W_1 = PW_2P^\dagger\quad\text{ and }\quad
    W_1^\dagger = PW_2^\dagger P^\dagger.
  \end{align*}
\end{lemma}

\begin{proof}
  Let us define $P$ as $P=\diag(S,P_1,\ldots,P_{n-1})$ with
  \begin{align*}
    P_l = (U_1U_2\ldots U_l)^\dagger S
  \end{align*}
  for $l=1,\ldots,n-1$. Then one easily checks that $W_1 =
  PW_2P^\dagger$ and $W_1^\dagger = PW_2^\dagger P^\dagger$.
\end{proof}

\noindent Digraphs of the matrix $W_2$, appearing in Lemma
\ref{thm:equiv_sum}, correspond to digraphs with several disconnected
components, each being either a loop or a string. Therefore, using
this lemma together with Theorem \ref{thm:solution} we obtain the
following classification:

\begin{theorem}\label{thm:equiv_rep}
  Let $\phi$ be a locally injective hermitian representation
  of $\Cnbg$.  Then $\phi$ is unitarily equivalent to a representation
  whose graph is such that every connected component is
  either a string or a loop.
\end{theorem}

\noindent Moreover, we can prove that a representation can be reduced
no further. 

\begin{proposition}
  Let $\phi$ be a hermitian representation of $\Cnbg$ written in the
  form of Theorem \ref{thm:solution}, and assume that the
  corresponding graph is a loop or a string. Then $\phi$ is irreducible.
\end{proposition}

\begin{proof}
  Let us start by proving the statement when $\phi$ is a loop
  representation.  For a representation in the form given by Theorem
  \ref{thm:solution}, and for which the corresponding graph is a loop,
  the dimension $N$ is the smallest positive integer such that
  $s^N(\xv_i)=\xv_i$ for all vertices $i$ in the loop. If a smaller
  such $N$ existed, then the corresponding graph would not be a loop,
  when brought to the form of Theorem \ref{thm:solution}. Now, assume
  that $\phi$ is reducible. Then there exists a representation $\phi'$
  of dimension $N'<N$, and by Lemma \ref{lemma:equiv_permutation} we
  know that
  $\{\xv'_1,\ldots,\xv'_{N'}\}\subseteq\{\xv_1,\ldots,\xv_N\}$. This
  implies that $\phi'$ is also a loop representation; hence, there
  must exist an integer $k\leq N'$ such that $s^{k}(\xv'_i)=\xv'_i$,
  which is impossible by the above argument. This proves that $\phi$
  is irreducible.

  Next, we assume that $\phi$ is a reducible string representation of
  dimension $N$. Then there exists a string representation $\phi'$ of
  dimension $N'<N$, which implies that there exists a string of length
  $k<N$. Since $\Spec(\phi')\subseteq\Spec(\phi)$, this contradicts
  the existence of the original string of length $N$. Hence, $\phi$ is
  irreducible.
\end{proof}

\noindent The next proposition tells us when two irreducible representations of
the same dimension are equivalent.

\begin{proposition}\label{proposition:equiv_irred}
  Let $\phi$ and $\phi'$ be $N$-dimensional irreducible hermitian
  representations of $\Cnbg$. Then $\phi$ and $\phi'$ are equivalent
  if and only if $\Spec(\phi)=\Spec(\phi')$ and
  $\det\phi(W)=\det\phi'(W)$.
\end{proposition}

\begin{proof}
  First, assume that $\phi$ is equivalent to $\phi'$. It follows
  directly that $\det\phi(W)=\det\phi'(W)$. Furthermore, it follows
  from Lemma \ref{lemma:equiv_permutation} that
  $\Spec(\phi)=\Spec(\phi')$. 

  Now, assume that $\Spec(\phi)=\Spec(\phi')$ and that
  $\det\phi(W)=\det\phi'(W)$.  Since the representations are
  irreducible, the spectrum consists of $N$ distinct vectors.
  Therefore, there exists a permutation $\sigma$ such that
  \begin{align*}
    \phi(D)=\sigma^\dagger\phi'(D)\sigma\quad\text{ and }\quad\phi(\Dt)=\sigma^\dagger\phi'(\Dt)\sigma.
  \end{align*}
  Let us define a representation $\psi$ by
  $\psi(W)=\sigma^\dagger\phi'(W)\sigma$; clearly, $\psi$ is
  equivalent to $\phi'$. Since 
  \begin{align*}
    |\phi(W)_{i,i+1}|^2=\phi(D)_{ii}=\psi(D)_{ii}=|\psi(W)_{i,i+1}|^2,    
  \end{align*}
  we have that $|\phi(W)_{ij}|=|\psi(W)_{ij}|$ for $i,j=1,2,\ldots,N$.
  Moreover, since $\det\phi(W)=\det\psi(W)$, there exists a diagonal
  unitary matrix $P$ such that
  \begin{equation*}
    \phi(W)=P^\dagger\psi(W)P=P^\dagger\sigma^\dagger\phi'(W)\sigma P.\qedhere
  \end{equation*}
\end{proof}

\noindent Note that for a string representation $\phi_S$,
it is always true that $\det\phi_S(W)=0$.

Let us present the matrices of irreducible representations and their corresponding digraphs.
The matrix of an irreducible loop representation has the form
\begin{equation*}
  W = 
  \begin{pmatrix}
    0 & W_{12} & 0     & \cdots & 0\\
    0 & 0     & W_{23} & \cdots & 0\\
    \vdots & \vdots & \ddots & \ddots & \vdots\\
    0 & 0 & \cdots & 0 & W_{N-1,N}\\
    W_{N,1} & 0 & \cdots & \cdots & 0
  \end{pmatrix}
\end{equation*}
and the corresponding digraph is

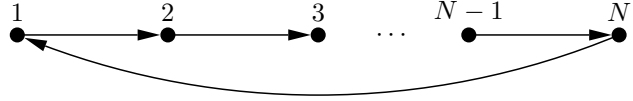
\begin{graph}(10,2)(-2,-1.5)
  \roundnode{n1}(0,0)
  \roundnode{n2}(2,0)
  \roundnode{n3}(4,0)
  \roundnode{n4}(6,0)
  \roundnode{n5}(8,0)
  \autonodetext{n1}[n]{1}
  \autonodetext{n2}[n]{2}
  \autonodetext{n3}[n]{3}
  \autonodetext{n4}[n]{$N-1$}
  \autonodetext{n5}[n]{$N$}
  \diredge{n1}{n2}
  \diredge{n2}{n3}
  \diredge{n4}{n5}
  \dirbow{n5}{n1}{0.1}
  \freetext(5,0){$\cdots$}[\opaquetextfalse]
\end{graph}

\noindent For an irreducible string representation we get
\begin{equation*}
  \raggedleft
  W = 
  \begin{pmatrix}
    0 & W_{12} & 0     & \cdots & 0\\
    0 & 0     & W_{23} & \cdots & 0\\
    \vdots & \vdots & \ddots & \ddots & \vdots\\
    0 & 0 & \cdots & 0 & W_{N-1,N}\\
    0 & 0 & \cdots & \cdots & 0
  \end{pmatrix}
\end{equation*}
with the following digraph

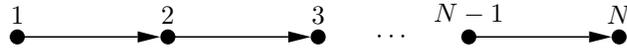
\begin{graph}(10,0.8)(-2,0)
  \roundnode{n1}(0,0)
  \roundnode{n2}(2,0)
  \roundnode{n3}(4,0)
  \roundnode{n4}(6,0)
  \roundnode{n5}(8,0)
  \autonodetext{n1}[n]{1}
  \autonodetext{n2}[n]{2}
  \autonodetext{n3}[n]{3}
  \autonodetext{n4}[n]{$N-1$}
  \autonodetext{n5}[n]{$N$}
  \diredge{n1}{n2}
  \diredge{n2}{n3}
  \diredge{n4}{n5}
  \freetext(5,0){$\cdots$}[\opaquetextfalse]
\end{graph}

\section{Constructing representations}

\noindent Let $\xv\in\RRplus=\{(x,y)\in\reals^2\st x>0,y>0\}$ be a
periodic point of period $N$, i.e. $s^N(\xv)=\xv$ but
$s^k(\xv)\neq\xv$ for $k=1,\ldots,N-1$, for the map
\begin{align*}
  s:\twovec{x}{y}\longrightarrow
  \twovec{\displaystyle \alpha+\sumkn\parab{\beta_ky^k+\gamma_k x^k}}{x},
\end{align*}
such that $s^k(\xv)\in\RRplus$ for $k=1,\ldots,N-1$. Writing
$s^{k}(\xv)=(d_k,\dt_k)$, it is easy to check that $W$ with
\begin{align*}
  &W_{k,k+1}=\sqrt{d_k}\qquad\text{ for }k=1,\ldots,N-1\\
  &W_{N,1}=e^{i\gamma}\sqrt{d_N}
\end{align*}
is an irreducible $N$-dimensional loop representation of $\Cnbg$ for
all $\gamma\in\reals$. Hence, by Proposition
\ref{proposition:equiv_irred}, every periodic orbit contained in
$\RRplus$, gives rise to a one-parameter family of inequivalent loop
representations, since
$\det W=e^{i\gamma}\sqrt{d_1d_2\cdots d_N}$. Moreover, disjoint orbits of
the same period correspond to inequivalent irreducible representations
of the same dimension.

In the same way, each $\xv=(a,0)$ with $a>0$ such that
$s^{N-1}(\xv)=(0,b)$, with $b>0$, and $s^k(\xv)\in\RRplus$ for
$k=1,\ldots,N-2$, gives an irreducible $N$ dimensional string representation of
$\Cnbg$. Namely, we define $W$ through
\begin{align*}
  W_{k,k+1} = \sqrt{d_k}
\end{align*}
with $(d_k,\dt_k)=s^{k-1}(\xv)$ for $k=1,\ldots,N-1$. Since $\det
W=0$, there is no parameter giving inequivalent representations. If we
define a $N$-string for $s$ to be such a set of points, each
$N$-string correspond to an irreducible string representation of
$\Cnbg$.

From these considerations we conclude that finding all irreducible
hermitian representations of $\Cnbg$ is equivalent to finding all
periodic orbits in $\RRplus$ and $N$-strings for the dynamical map.

\section{Representations of $\Cbg{1}$}

\noindent In the case when the algebra is of order one, the dynamical
map will be an affine map from the plane to itself. This allows us to
work out the representations quite explicitly. In particular, $s$ is
invertible which implies that every representation is locally
injective. Hence, we can classify representations according to Theorem
\ref{thm:equiv_rep}.

The defining relations for $\Cbg{1}$ are\footnote{Note that similar equations are under consideration in \cite{HopLee}.}
\begin{align*}
  &W^2V=\alpha W+\beta_1VW^2+\gamma_1WVW\\
  &WV^2=\alpha V + \beta_1 V^2W+\gamma_1 VWV
\end{align*}
or, in terms of $X,Y,Z$
\begin{align*}
  [X,Y]&=Z\\
  [Y,Z]&=\frac{\alpha}{2}X + \half(\beta_1+\gamma_1-1)X^3+\half(\beta_1-\gamma_1+3)YXY\\
  &\qquad\qquad+ \halfi(\beta_1+1)\para{X^2Y-YX^2}+\half(\gamma_1-2)\para{XY^2+Y^2X}\\
  [Z,X]&=\frac{\alpha}{2}Y + \half(\beta_1+\gamma_1-1)Y^3 + \half(\beta_1-\gamma_1+3)XYX\\
  &\qquad\quad + \halfi(\beta_1+1)\para{XY^2-Y^2X}+\half(\gamma_1-2)\para{YX^2+X^2Y}.
\end{align*}
For convenience, let us write $\beta_1=q-p^2$ and $\gamma_1=2p$.
Finding irreducible hermitian representations of this algebra
corresponds to finding periodic orbits and $N$-strings for the affine map
\begin{align*}
  s:\twovec{x}{y}\longrightarrow\twomatrix{2p}{q-p^2}{1}{0}\twovec{x}{y}+\twovec{\alpha}{0}.
\end{align*}
In the following, let us assume that $q<0$ for simplicity. The
representations for these algebras were found and studied in
\cite{abhhs}, but let us recall some basic facts. We introduce the
following notation
\begin{align*}
  s(\xv)=A\xv+\cv\equiv \twomatrix{2p}{q-p^2}{1}{0}\twovec{x}{y}+\twovec{\alpha}{0}.
\end{align*}
It follows from the fact that $q<0$ that this map will always have a
unique fix-point $\xv_f$, which implies that $s$ amounts to a
\emph{linear} map around $\xv_f$. Let us study the periodic orbits of
$s$. A necessary condition for a periodic orbit in $\RRplus$ to exists
is that the eigenvalues $\mu,\lambda$ of $A$ satisfy
$\lambda^n=\mu^n=1$, which restricts the existence of loop
representations to certain algebras. Let us find a way to parametrize
them.  The (complex) eigenvalues of $A$ are
\begin{align*}
  &\lambda = p+\sqrt{q}\\
  &\mu = p-\sqrt{q},
\end{align*}
and demanding $|\lambda|=|\mu|=1$ gives us a parametrization through
\begin{align*}
  &p = \cos 2\theta\\
  &q = -\sin^2 2\theta
\end{align*}
with $0<\theta<\pi/2$. Furthermore, $\lambda^n=\mu^n=1$ implies that
$e^{i2n\theta}=1$, which gives $\theta=k\pi/n$ for some
$k\in\integers$.  Requiring that $n$ is the least period of the orbit
gives $\gcd(k,n)=1$ and if $\alpha>0$ we can always find periodic
orbits in $\RRplus$. Hence, for these algebras there are only
irreducible loop representations of dimension $n$. As we will see in
the next part, this fact is changed for higher order algebras, where
irreducible representations of all dimension might exist.

\section{Representations of a second order \Henon algebra}

\noindent The first order \Henon algebras are just the algebras
presented in the previous section. For these algebras, everything can
be explicitly calculated since the map $s$ is an affine map. When we
turn to the next order \Henon algebras, things become much more
involved. Nevertheless, the dynamical map is still invertible, which implies
that all representations can be decomposed into irreducible loop and
string representations. We will now continue to construct a second
order \Henon algebra for which irreducible loop representations of all
dimensions exist. Consider the \Henon map
\begin{align*}
  f:\twovec{x}{y}\longrightarrow\twovec{a-by-x^2}{x}
\end{align*}
For certain values of $a$ and $b$, e.g. $a=5$ and $b=0.3$ there is a
bounded subset $\Lambda\subset\reals^2$ such that $f|_\Lambda$ is
topologically conjugate to a two-sided shift on two symbols (see, e.g. \cite{dynsys}).
By definition, this means that there exists a bijection
$h:\Lambda\to\Sigma_2$, where
$\Sigma_2=\{(\ldots,a_{-1},a_0,a_1,\ldots)\st a_k\in\{0,1\}\text{ for
  all }k\in\integers\}$, and a shift map $\sigma:\Sigma_2\to\Sigma_2$
such that
$\sigma\paraa{(\ldots,a_{-1},a_0,a_1,\ldots)}=(\ldots,a_{0},a_1,a_2,\ldots)$
and $\sigma\circ h= h\circ f$. Thus, the periodic orbits of
$f|_\Lambda$ is in one-to-one correspondence with the periodic orbits
of $\sigma$ on $\Sigma_2$. Moreover, for the shift map on $\Sigma_2$
there exists periodic points of all periods. However, for our
purposes, we need periodic orbits contained in $\RRplus$, which is the
case if $\Lambda\subset\RRplus$.  This is not true for the map $f$,
but we can easily create a new map with this property.  Since
$\Lambda$ is a bounded set, there exists a positive number $r$ (for
$a=5$ and $b=0.3$ we can take $r>2.5$) such that
$\Lambda\subset\{(x,y)\in\reals^2\st |x|<r\text{ and }|y|<r)\}$. If we
define
\begin{align*}
  s:\twovec{x}{y}\longrightarrow\twovec{a+r-b(y-r)-(x-r)^2}{x}
\end{align*}
one can easily check that $f^n(\xv)+(r,r)=s^n\paraa{\xv+(r,r)}$. This
implies that every periodic orbit of $f$ in $\Lambda$ will appear as
a periodic orbit for $s$, contained in $\RRplus$. Hence, the second order \Henon algebra defined by
\begin{align*}
  &W^2V=(a+r+br-r^2)W-bVW^2+2rWVW-(WV)^2W\\
  &WV^2 =(a+r+br-r^2)V - bV^2W +2rVWV-V(WV)^2
\end{align*}
has irreducible loop representations of all dimensions for certain values of $a$, $b$ and $r$.

\end{document}